\documentclass[aps,pra,twocolumn,showpacs,superscriptaddress]{revtex4-1}

\usepackage{graphicx}
\usepackage{amsfonts,mathrsfs,amsmath,amsthm,amssymb,bbold}
\usepackage{mathrsfs,amsmath,amssymb,bbold}
\usepackage{bm,mathrsfs}

\begin{document}

\title{Continuous monitoring measured signals bounded by past and future conditions in enlarged quantum systems}


\author{Le Bin Ho}
\thanks{Electronic address: binho@kindai.ac.jp}
\affiliation{Department of Physics, Kindai University, Higashi-Osaka, 577-8502, Japan,}
\affiliation{Ho Chi Minh City Institute of Physics, VAST, Ho Chi Minh City, Viet Nam}

\begin{abstract}
In a quantum system that is bounded by past and future conditions, 
weak continuous monitoring forward-evolving 
and backward-evolving quantum states 
are usually carried out separately. 
Therefore, measured signals at a given time $t$ 
cannot be monitored continuously. 
Here, we propose an {\it enlarged-quantum-system} 
method to combine these two processes together. 
Therein, we introduce an enlarged quantum state 
that contains both the forward- and backward-evolving quantum states.  
The enlarged state is governed by an enlarged master equation 
and propagates one-way forward in time. 
As a result, the measured signals at time $t$ can be monitored continuously
and can provide advantages in the signals amplification and signal processing techniques.
Our proposal can be implemented on various physical systems, 
such as superconducting circuits, NMR systems, ion-traps, 
quantum photonics, and among others.
\end{abstract}

 \pacs{03.65.Aa, 03.65.Ca, 03.65.Ta}

\maketitle

\section{Introduction}
\label{seci}

In quantum mechanics, measurement results 
at a given time can be predicted from 
the system of interest that propagates forward in time 
from past conditions. Instead, if the system is bounded by 
future conditions and propagates backward in time, 
then the results are exactly the time reversal of the former case 
when the past and the future conditions are prepared in the same state \cite{Campagne112,Foroozani116,Tan114}. 
This is the time-reversal symmetry in quantum mechanics 
(see Ref. \cite{Dressel119} for reconstructing the time-reversal symmetry.) 
Nonetheless, weak measurements that are bounded by both 
the past and the future conditions can affect the statistical results 
and provide more information about the measured system \cite{Aha60,Aha63}. 
In this case, a time-dependent weak value can be defined by incorporating both 
a {\it forward-evolving state} and a {\it backward-evolving state} 
at the same given time \cite{Wu83,Nakamura85}. 
However, it neither be obtained directly nor be monitored continuously 
because the quantum trajectories of the forward- 
and backward-evolving states are obtained separately \cite{Campagne112,Foroozani116,Tan114,Dressel119,Tan96,Luis96,Murch502}. 
It implies that the measured system does not evolve 
causally from the past to the future. This is a {\it noncausal} problem 
in the measurements based on the two-state-vector formalism \cite{Ho97}.
Monitoring a system continuously over time can help to characterize 
the stochastic dynamics of the system during the measurement and
also might useful for quantum state reconstruction and parameter estimation theory 
\cite{Silberfarb95}.
Therefore, it is also beneficial and demand to monitor 
time-dependent weak values continuously.

In this paper, to solve the noncausal problem and obtain the continuous monitoring,
we propose an {\it enlarged-quantum-system} method, 
therein we map both the forward- and the backward-evolving states 
onto an enlarged quantum state. The enlarged quantum state can 
propagate casually one-way forward in time that is governed by 
an enlarged master equation. We also introduce a 
{\it two-time correlation weak value}, where we show that 
it can be monitored continuously in the enlarged system. 
We illustrate our proposal in a superconducting qubit 
driven at resonance based on the experiments in 
Refs. \cite{Campagne112,Foroozani116,Tan114}. 
Afterward, we also discuss how to implement 
the enlarged system in various physical platforms, 
such as superconducting circuits, NMR systems, 
ion-traps, and quantum photonics systems. 

The structure of this paper is organized as follows. 
Section \ref{secii} introduces an enlarged system
where both the forward- and backward-evolving states 
are embedded onto an enlarged state. The master equation
that governs the evolution of the enlarged state is also discussed.
We introduce the two-time correlation 
weak value in Sec. \ref{seciii} and illustrate it in Sec. \ref{seciv}. 
In Sec. \ref{secv}, we discuss the implementation of the enlarged system. 
The paper concludes with a discussion and a brief summary in Sec. \ref{secvi}.

\section{Enlarged quantum system}\label{secii}
\subsection{Enlarged quantum state}

We first describe a mapping process that maps two arbitrary states, 
such as $\rho$ and $E$ in an original system ($\mathcal{OS}$) onto 
an enlarged state $\varrho$ in an enlarged system ($\mathcal{ES}$).
In the $\mathcal{OS}$, the complex Hilbert space of 
a $d$-dimensional vector is denoted as $\mathbb{C}_d$ and
the complex Hilbert space of a $d\times d$ density matrix is set to be 
$L(\mathbb{C}_d)$.
We consider a mapping process from the original Hilbert space 
$L(\mathbb{C}_d)$ to an enlarged Hilbert space 
$L(\mathbb{C}_2\otimes \mathbb{C}_d)$ that maps both 
$\rho$ and $E$ onto $\varrho$ in the following 
\begin{align}{\label{varrho}} 
\varrho_{t} = \dfrac{1}{2}
	\begin{pmatrix}
  \rho_\tau & 0_d \\
  0_d & E_{\tau'} 
 \end{pmatrix}\;,
\end{align}
where $0_d \in L(\mathbb{C}_d)$ is a $d\times d$ zero matrix. 
Factor $\frac{1}{2}$ is used for the normalization. 
In the following subsection, we will choose $\rho$ as a forward-evolving state
and $E$ as a backward-evolving state.
For now, however, we treat them in general forms.
We note that $t,\ \tau,$ and $\tau'$ are different, in general.
A similar mapping process for pure quantum states 
has been introduced previously \cite{Ho97}. 
This mapping can be implemented by adding an ancillary qubit 
to the $\mathcal{OS}$ such that 
$\varrho_t = [ |0\rangle\langle 0|\otimes\rho_\tau
+ |1\rangle\langle 1|\otimes E_{\tau'}]/2$, 
where $|0\rangle \equiv {{1}\choose{0}}$ and  
$|1\rangle \equiv {{0}\choose{1}}$ are the bases of 
the ancillary qubit. Recently, similar mapping has been extensively studied 
both in theoretical and experimental 
\cite{Casanova1,Noh87,Lara89,Zhang6,Candia111,Pedernales90,Loredo116,Chen116,Rodriguez111,Ho383}. 
The states in the $\mathcal{OS}$ can be decoded by the inversions
\begin{align}{\label{decord}}
\rho_\tau = 2\mathcal{M}\varrho_{t}\mathcal{N} \ \text{and} 
\ E_{\tau'} = 2\mathcal{M}\varrho_{t}(\sigma_x\otimes\bm{I}_n)\mathcal{N},
\end{align}
where $\mathcal{{M}}=(1,1)\otimes \bm{I}_d$ and 
$\mathcal{N}={{1}\choose{0}}\otimes \bm{I}_d$, 
where $\bm{I}_d\in L(\mathbb{C}_d)$ is a $d\times d$ identity matrix. 

\subsection{Enlarged master equation}
Now we describe the master equation in the $\mathcal{ES}$. 
We consider the case that the $\mathcal{OS}$ is bounded by 
a past condition $\rho_0$ and a future condition $E_T$ for a time interval $[0, T]$.
The forward-evolving state $\rho_t$ satisfies the Lindblad master equation 
\cite{Wiseman,Jacobs47}
\begin{align}{\label{Lindblad_forward}}
\dfrac{d\rho_t}{dt} = -\dfrac{i}{\hbar}[\bm{H}, \rho_t] 
+\sum_n\dfrac{1}{2}\Bigl[2\bm{C}_n\rho_t\bm{C}_n^\dagger 
- \{\bm{C}_n^\dagger \bm{C}_n,\rho_t\}\Bigr],
\end{align}
which propagates forward in time from $t = 0$ to $t$, 
where  $\bm{H}\in L(\mathbb{C}_d)$ is the Hamiltonian of the $\mathcal{OS}$, 
$\bm{C}_n = \sqrt{k_n}\bm{A}_n$ is a Lindblad operator 
 $\in L(\mathbb{C}_d)$,
that describes the effect of the environment 
in the Markov approximation, and $\bm{A}_n$ is an operator 
through which the environment couples to the system with a relaxation rate $k_n$.
Similarly, the Lindblad master equation, 
which governs the evolution of the backward-evolving state $E$, 
is given by \cite{Campagne112,Foroozani116,Tan114,Gamm111}
\begin{align}\label{Lindblad_back}
\dfrac{dE_t}{dt} = -\dfrac{i}{\hbar}[\bm{H}, E_t] 
-\sum_n\dfrac{1}{2}\Bigl[2\bm{C}_n^\dagger E_t\bm{C}_n 
- \{\bm{C}_n^\dagger \bm{C}_n, E_t\}\Bigr],
\end{align}
which propagates backward in time from $T$ to $t\leq T$.
We note that in the time interval $t \in [0, T]$, 
this backward evolution has a {\it forward version} 
that evolves forward in time and satisfies 
the time-reversal-symmetry property \cite{Dressel119}. 
More precisely, the quantum trajectory of the backward evolution, $E_t$,
has a time-reversal trajectory, $E_{T-t}$. 
(See Appendix \ref{appA}). 
The forward version is
\begin{align}\label{Lindblad_back_for}
\notag\dfrac{dE_{T-t}}{dt} &= \dfrac{i}{\hbar}[\bm{H}, E_{T-t}] \\
&+\sum_n\dfrac{1}{2}\Bigl[2\bm{C}_n^\dagger E_{T-t}\bm{C}_n 
- \{\bm{C}_n^\dagger \bm{C}_n, E_{T-t}\}\Bigr],
\end{align}
where its solution at time $t$ is $E_{T-t}$. 

Notable that Eqs. (\ref{Lindblad_forward}, 
\ref{Lindblad_back_for}) evolve forward in time.
We, therefore, combine them into an enlarged master equation 
which governs the enlarged state one-way forward in time as follows
\begin{align}\label{Lindblad_enlar}
\dfrac{d\varrho_{t}}{dt} = -\dfrac{i}{\hbar}[\bm{\mathcal{H}}, \varrho_{t}] 
+\sum_n\dfrac{1}{2}\Bigl[2\bm{\mathcal{C}}_n\varrho_{t}\bm{\mathcal{C}}_n^\dagger
 - \{\bm{\bar\mathcal{C}}_n^\dagger \bm{\bar\mathcal{C}}_n, \varrho_{t}\}\Bigr],
\end{align}
where we have defined the enlarged Hamiltonian 
$\bm{\mathcal{H}} \equiv\sigma_z\otimes \bm{H}$, 
the Lindblad operator $\bm{\mathcal{C}} \equiv 
|0\rangle\langle 0|\otimes \bm{C}+|1\rangle\langle 1|
\otimes \bm{C}^\dagger$, and $\bm{\bar\mathcal{C}}
\equiv \bm{I}_2\otimes \bm{C}$ in the $L(\mathbb{C}_2
\otimes\mathbb{C}_d)$ enlarged Hilbert space. 
The solution $\varrho_{t}$ at time $t$ is given by 
$\varrho_{t}$ in Eq. \eqref{varrho} where 
\begin{align}{\label{varrho_1}} 
\varrho_{t} = \dfrac{1}{2}
	\begin{pmatrix}
  \rho_t & 0_d \\
  0_d & E_{T-t} 
 \end{pmatrix}\;.
\end{align}
The enlarged trajectory (described by $\varrho_{t}$)
can be measured continuously by monitoring the $\mathcal{ES}$
forwardly in time.

We emphasize that the enlarged state $\varrho_{t}$ is
different from  the ``two-state" or ``density state" defined by 
Reznik and Aharonov \cite{Reznik52}
and later used by \cite{Shikano43,Oreshkov18,Silva19}.
In their original proposal, the density state is formed by putting the
pre- and postselected states in such as way that 
$\wp_t \equiv |\psi_t\rangle\langle\phi_t|$,
where $\wp_t$ is the density state, $|\psi_t\rangle$
and $|\phi_t\rangle$ are pre- and postselected states, respectively. 
Recently, Vaidman et al. \cite{Vaidman96}
also defined a so-called ``genuine mixed two-state vector" 
where $\wp_t \equiv (E_t,\rho_t)$.
Event in this case, the mapping is also different from ours:
While Vaidman's mapping is $L(\mathbb{C}_d) 
\to \mathbb{C}_2\otimes L(\mathbb{C}_d)$, our mapping is
$L(\mathbb{C}_d) \to L(\mathbb{C}_2\otimes \mathbb{C}_d)$.
Moreover, as we can see from Eq. \eqref{varrho_1}, $\rho_t$ and $E_{T-t}$
are different in time, while previous studies require a consistent time.

\section{Two-time correlation weak value}\label{seciii}

\subsection{Conventional time-dependent weak value}
In a conventional weak measurement,
the conventional time-dependent weak value is described by both the forward-evolving state
$\rho_t$ immediately before the weak measurement was carried out and the 
backward-evolving state $E_t$ immediately after the measurement \cite{Aha60,Aha72}. 
The conventional time-dependent weak value for an observable 
$\bm{A}$ at time $t$ is given by \cite{Wu83,Nakamura85,Vaidman96}
\begin{align}\label{tweak_value_enlar}
 \langle\bm{A}_t\rangle_{\rm {w}} = 
 \dfrac{{\rm Tr}[E_t\bm{A}\rho_t]}{{\rm Tr} [E_t\rho_t]},
\end{align}
where the subscript w stands for ``weak value." 

We now describe the conventional time-dependent weak value
in the $\mathcal{ES}$.
From Eq. \eqref{decord} we have
\begin{align}{\label{decord_1}}
\rho_t = 2\mathcal{M}\varrho_{t}\mathcal{N} \ \text{and} 
\ E_{\tau'} = 2\mathcal{M}\varrho_{t}(\sigma_x\otimes\bm{I}_n)\mathcal{N}.
\end{align}
Using the time-reversal evolution, we can derive 
$E_t = 2\mathcal{M}\varrho_{T-t}(\sigma_x\otimes\bm{I}_n)\mathcal{N}$. 
Substituting $\rho_t$ and $E_t$ to Eq. \eqref{tweak_value_enlar} we obtain
\begin{align}\label{weak_value_enlar}
 \langle\bm{A}_t\rangle_{\rm {w}} =\dfrac{{\rm Tr}
 [ \mathcal{M}\varrho_{T-t}(\sigma_x\otimes\bm{I}_d)
 \mathcal{N}\bm{A}\mathcal{M}\varrho_t\mathcal{N}]}{{\rm Tr} 
 [ \mathcal{M}\varrho_{T-t}(\sigma_x\otimes\bm{I}_d)
 \mathcal{N}\mathcal{M}\varrho_t\mathcal{N}]}.
\end{align}
Clearly, $\langle\bm{A}_t\rangle_{\rm {w}}$
depends on both $\varrho_{t}$ and $\varrho_{T-t}$. 
Therefore, it cannot be measured continuously in time 
even for the $\mathcal{ES}$ case.

\subsection{Two-time correlation weak value}
\begin{figure} [t!]
\centering
\includegraphics[width=8cm]{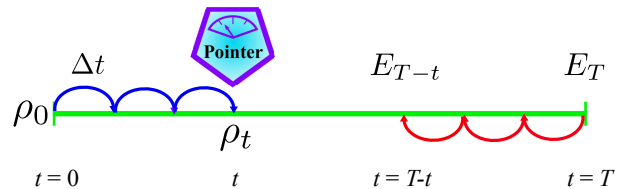}
\caption{
(Color online) Graphical scheme for two-time correlation weak values.
A quantum system is prepared in $\rho_0$ and post-selected onto $E_T$. 
The initial state propagates forward in time to $t$, 
which we name as the forward-evolving state $\rho_t$. 
The final state propagates backward from $T$ to $T-t$,
which we denote as the backward-evolving state $E_{T-t}$.
We also assume there are no further measurements after time $T$. 
For continuous measurements, the pointer is shifted (in time scale)
after each time interval $\Delta t$.
We emphasize that this measurement scheme is different from the original motivation 
by Aharonov et al., who consider a conventional weak value at time $t$ 
described by $\rho_t$ and $E_t$ \cite{Aha60}.
Even though this scheme cannot realize in an $\mathcal{OS}$, 
its two-time correlation weak value can be obtained continuously 
in an $\mathcal{ES}$ as we show in the main text.
}
\label{fig1}
\end{figure}

To enjoy the benefit of the $\mathcal{ES}$, we 
introduce a so-called {\it two-time correlation} 
weak value in a very similar way that
\begin{align}\label{t2weak_value_enlar}
 \langle\bm{A}_{t,T-t}\rangle_{\rm {w}}^{\rm c} = 
 \dfrac{{\rm Tr}[E_{T-t}\bm{A}\rho_t]}{{\rm Tr} [E_{T-t}\rho_t]},
\end{align}
where the superscript c represents ``two-time correlation."
Following are some properties of the two-time correlation
weak value.\\
(\textbf{i}) It is different from the conventional weak value: 
while the conventional weak value is conditioned on 
$\rho_t$ and $E_{t}$, the two-time correlation
weak value is described by $\rho_t$ and $E_{T-t}$,
as we depict in Fig. \ref{fig1}.
Obviously, it depends on
two different times, which are correlated, i.e., $t$ and $T-t$.
It coincides with the conventional weak value 
only at $t = T/2$, i.e., $t = T-t$.\\
(\textbf{ii})
The two-time correlation weak value defined by
Eq. \eqref{t2weak_value_enlar} is a mathematical concept
and thus, cannot be realized in the $\mathcal{OS}$.\\
(\textbf{iii})
However, in the $\mathcal{ES}$, 
we point out that the two-time correlation weak value is
an expectation value which can be measured continuously in time.
Substituting Eq. \eqref{decord_1} to Eq. \eqref{t2weak_value_enlar}
we obtain
\begin{align}\label{2weak_value_enlar}
 \langle\bm{A}_{t,T-t}\rangle^{\rm c}_{\rm {w}} =
 \dfrac{{\rm Tr}[ \mathcal{M}\varrho_t(\sigma_x\otimes\bm{I}_d)
 \mathcal{N}\bm{A}\mathcal{M}\varrho_t\mathcal{N}]}{{\rm Tr} 
 [\mathcal{M}\varrho_t(\sigma_x\otimes\bm{I}_d)\mathcal{N}
 \mathcal{M}\varrho_t\mathcal{N}]}
 =  \dfrac{{\rm Tr}[\bm{A}\tilde\varrho_t]}
 {{\rm Tr} 
 [\tilde\varrho_t]},
\end{align}
where $\tilde\varrho_t\equiv \mathcal{M}\varrho_t\mathcal{N}
\mathcal{M}\varrho_t(\sigma_x\otimes\bm{I}_d)\mathcal{N}$.
In this form, the two-time correlation weak value depends only on
$\varrho_t$ at time $t$.
Therefore, it can be monitored continuously at each time 
$t$ from 0 to $T$ causally.
In comparison to Eq. \eqref{weak_value_enlar}, 
the two-time correlation weak value Eq. \eqref{2weak_value_enlar}
is more promising for continuous monitoring signals.\\
(\textbf{iv})
Furthermore, two-time correlation weak values are also useful for
signals amplification and signals processing. 
It can be seen that the two-time correlation weak value can
excess outside the normal range of the observable eigenvalues
with a proper choice of the pre- and post-selected density states $\rho_0$
and $E_T$. See our illustration in Figs. \ref{fig3}, \ref{fig4} below for 
the continuous monitoring signals, signals amplification, and signals processing.

It is also worthwhile to note that Aharonov et al. 
have discussed the concept of 
multiple-time states and multiple-time measurements \cite{Aha79}.
However, we note that it is different from our ``two-time correlation" here.
In their work, they consider multiple preparations and measurements, 
such that 
{\it preparation $\to$ measurement $\to$ preparation $\to$ measurement $\to$ ...}
For a ``two-time state," in their words, it means 
{\it preparation $\to$ measurement $\to$ preparation},
or in other words, it implies 
{\it preparation $\to$ measurement $\to$ postselection}. 
In our work here, we consider only this situation 
and require no further measurements after the postselection.
Their ``two-time state" means time in the preparation and time in the postselection.
Whereas, by ``two-time correlation" in this work, 
it means two times in between the preparation time
and postselection time.
Furthermore, in our work, we consider such two-time correlation 
weak values in an $\mathcal{ES}$ while the previous study did not.

\section{Illustration}\label{seciv}

To illustrate our proposal for some physical problems, 
we first consider an example based on a superconducting qubit 
driven at resonance as experimentally studied in Ref. \cite{Campagne112}, 
where the qubit is coupled to a waveguide cavity 
\cite{Koch76,Paik107,Vijay490,Weber511}.
So far, weak measurements under the presence of decoherence  
have been investigated but they focused only on the $\mathcal{OS}$, where 
the continuous monitoring is not discussed \cite{Shikano43,Abe2}.
Here, we consider such problem in the $\mathcal{ES}$ 
with our two-time correlation weak values.
 We will analyze the conventional weak value 
 and two-time correlation weak value 
 of the fluorescence signal, the atom population, 
 and the photon number in some concrete models. 
We show that in the case of two-time correlation weak value,
these measured signals can be detected continuously 
 in the $\mathcal{ES}$.

A specific model of a two-level atom, which is driven by a laser field 
at the Rabi oscillations, oscillates between the ground state 
$|g\rangle$ and the excited state $|e\rangle$. 
These oscillations emit a so-called fluorescence signal, 
which is detected due to the transition from the excited state 
to the ground state. Its amplitude is proportional to 
the average value of the lowering operator 
$\langle\sigma_-\rangle$ \cite{Campagne112}. 
In the rotating wave approximation, we write the laser Hamiltonian as 
$\bm{H}_L = \hbar\Omega\sigma_y/2$,  where $\Omega$ is the Rabi frequency. 
The Lindblad operator is $\bm{C} = \sqrt{k}\sigma_-$. 
In this model, the forward master equation which governs 
the forward-evolving state $\rho$ is given by
\begin{align}
\dfrac{d\rho_t}{dt} &= -\dfrac{i\Omega}{2}\Bigl[\sigma_y, \rho_t\Bigr] 
+k\Bigl[\sigma_-\rho_t\sigma_+ - \dfrac{1}{2}\{\sigma_+ 
\sigma_-,\rho_t\}\Bigr]. \label{Lindblad_for}
\end{align}
The backward-evolving state $E$ 
is governed backward in time by a corresponding adjoint equation as 
\begin{align}
\dfrac{dE_t}{dt} &= -\dfrac{i\Omega}{2}\Bigl[\sigma_y, E_t\Bigr] 
-k\Bigl[\sigma_+E_t\sigma_- - \dfrac{1}{2}\{\sigma_+ \sigma_-,E_t\}\Bigr], 
\label{Lindbladr_back}
\end{align}
where we have used the standard Pauli matrices 
$\sigma_z = |e\rangle\langle e|- |g\rangle\langle g|$
 and $\sigma_y=i(\sigma_--\sigma_+)$.
The enlarged master equation 
Eq. \eqref{Lindblad_enlar} takes the form
\begin{align}\label{Lindblad_enlar_exam1}
\dfrac{d\varrho_t}{dt} = -\dfrac{i\Omega}{2}
\Bigl[\sigma_z\otimes \sigma_y, \varrho_t\Bigr] 
+k\Bigl[\bm{\mathcal{C}}\varrho_t\bm{\mathcal{C}}^\dagger 
- \dfrac{1}{2}\{\bm{\bar\mathcal{C}}^\dagger \bm{\bar\mathcal{C}}, \varrho_t\}\Bigr],
\end{align}
where $\bm{\mathcal{C}} = 
|0\rangle\langle 0|\otimes\sigma_-+|1\rangle\langle 1|
\otimes \sigma_+$ and $\bm{\bar\mathcal{C}} = 
\bm{I}_2\otimes\sigma_-$. 
Solving this enlarged equation will give 
the enlarged state $\varrho_t$ at any time $t \in [0,T]$.

For concreteness, we choose the parameters as
 $\Omega/2\pi = 1.16$ MHz, $k/2\pi = 95$ kHz 
 \cite{Foroozani116}. The past condition at time $t=0$ is 
 $\rho_0=|g\rangle\langle g|$ and the future condition at time 
 $T$ is $E_T = |g\rangle\langle g|$. We next examine the measured signals: 
 the atom population $\langle\sigma_z\rangle$, 
 the photon number $\langle n\rangle$, 
 and the fluorescence signal $\langle\sigma_-\rangle$.

A measured signal can be detected by 
continuously monitoring the cavity, 
which can be described by the theory of POVM. 
For example, the measurement of the voltage signal $V$, 
that describes the atom population $\langle\sigma_z\rangle$, 
is given by the POVM operator as \cite{Foroozani116,Tan114}
\begin{align}\label{Omg_V}
\Omega_V = (2\pi a^2)^{-1/4}e^{-(V-\sigma_z)^2/4a^2}.
\end{align}
The probability of the outcome $V$ 
that depends only on the enlarged state is shown in 
Appendix \ref{appB}, where
\begin{align}\label{PV}
\notag P(V) &= \dfrac{{\rm Tr}
(\Omega_V\rho_t\Omega_V^\dagger E_t)}
{\sum_V {\rm Tr}(\Omega_V\rho_t\Omega_V^\dagger E_t)}\\
\notag& \propto \varrho^{00}_t\varrho^{22}_{T-t}
e^{-(V-1)^2/2a^2}+\varrho^{11}_t\varrho^{33}_{T-t}e^{-(V+1)^2/2a^2}\\
& +(\varrho^{10}_t\varrho^{23}_{T-t}+\varrho^{01}_t\varrho^{32}_{T-t})e^{-(V^2+1)/2a^2},
\end{align}
where $\varrho^{ij}$ are the elements of the enlarged density matrix.
The conventional weak value of the voltage signal is given by 
$\langle V\rangle_{\rm w} = \int P(V)V dV$, and can be evaluated
\begin{align}\label{mean_V}
\langle V\rangle_{\rm w} =  \dfrac{\varrho^{00}_t\varrho^{22}_{T-t}
-\varrho^{11}_t\varrho^{33}_{T-t}}{\varrho^{00}_t\varrho^{22}_{T-t}
+\varrho^{11}_t\varrho^{33}_{T-t}+\varrho^{10}_t\varrho^{23}_{T-t}
+\varrho^{01}_t\varrho^{32}_{T-t}}.
\end{align}
For the two-time correlation weak value, it yields
\begin{align}\label{mean_2V}
\langle V\rangle^{\rm c}_{\rm w} =\dfrac{\varrho^{00}_t\varrho^{22}_t
-\varrho^{11}_t\varrho^{33}_t}{\varrho^{00}_t\varrho^{22}_t
+\varrho^{11}_t\varrho^{33}_t+\varrho^{10}_t\varrho^{23}_t
+\varrho^{01}_t\varrho^{32}_t}.
\end{align}
Note that in this section we omit $t$ and $T-t$ in the 
conventional weak value and two-time weak value.
\begin{figure} [t]
\centering
\includegraphics[width=8.6cm]{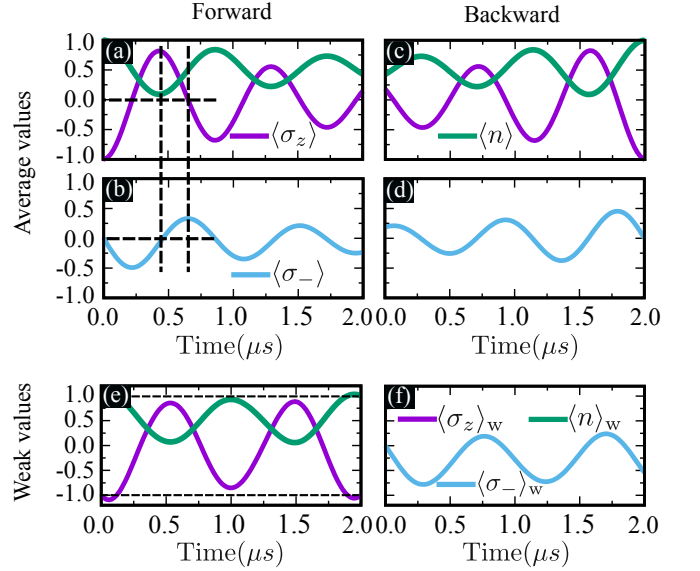}
\caption{
(Color online) The average values and conventional weak values 
of the measured signals: the atom population 
$\langle\sigma_z\rangle$ (dark-violet curves), 
the photon number $\langle n\rangle$ (vine-green curves), 
and the fluorescence signal $\langle\sigma_-\rangle$ (soft-blue) 
in a two-level atom. (a, b): The measured values correspond to 
the system that prepared in the density state 
$\rho_0 = |g\rangle\langle g|$ at time $t=0$. 
The forward-evolving state at time $t$ is given by 
Eq. \eqref{Lindblad_for} and propagates forward in time. 
(c) and (d) show the same measured values correspond to 
the system that postselected onto the density state 
$E_T=|g\rangle\langle g|$ at time $T$. 
The backward-evolving state at time $t$ is given by 
Eq. \eqref{Lindbladr_back} and propagates backward in time. 
(e) and (f) represent the conventional weak values of the atom population 
$\langle\sigma_z\rangle_{\rm w}$, the photon number 
$\langle n\rangle_{\rm w}$, and the fluorescence signal 
$\langle\sigma_-\rangle_{\rm w}$ correspond to 
the system prepared in the state $\rho_0$ at time $t=0$ 
and postselected onto the state $E_T$ at time $T$. 
These conventional weak values can be obtained by both methods 
as shown in Eqs. (\ref{tweak_value_enlar}, \ref{weak_value_enlar}), 
where the enlarged quantum state at time $t$ 
is given by Eq. \eqref{Lindblad_enlar_exam1} 
and propagates forward in time.
}
\label{fig2}
\end{figure}

Fig. \ref{fig2}(a) shows the average expectation values 
of the atom population and the photon number 
conditioned on the forward-evolving state, 
e.g., $\langle\sigma_z\rangle = {\rm Tr}[\rho_t\sigma_z]$. 
The results show the gradual dephasing of the atom 
due to the interaction \cite{Campagne112,Foroozani116}. 
The atom population is bounded in the interval $[-1,+1]$. 
It is a $\pi$-phase difference from the photon number 
which implies that the atom absorbs photons to 
transfer from the ground state $|g\rangle$ to 
the excited state $|e\rangle$ and vice versa. 
In Fig. \ref{fig2}(b), the fluorescence signal, 
which is also conditioned on the forward-evolving state, 
is the $\pi/2$-phase difference from the atom population. 
In detail, starting from the maximum atom population 
(excited state), the fluorescence signal is zero. 
Then, during the relaxation from the maximum to 
zero of the atom population, the fluorescence signal 
increases and reaches the maximum as shown by 
the dash lines in Fig. \ref{fig2}(a, b). 
The process keeps going afterward. 
Similarly, Fig. \ref{fig2}(c, d) examine the average values of 
the measured signals conditioned on the backward-evolving state, 
e.g., $\langle\sigma_z\rangle = {\rm Tr}[E_t\sigma_z]$. 
In this case, the atom population and the photon number 
are damped backward in time and equal to the time reversal 
of the measured signals in Fig. \ref{fig2}(a) 
while the fluorescence signal reverts both the time and sign 
in comparison to the former case Fig. \ref{fig2}(b). 
The conventional weak values of these measured signals conditioned on 
both the forward- and backward-evolving states 
are shown in Fig. \ref{fig2}(e, f). Notable, the results do not ``damp" 
and can excess beyond their normal intervals due to the interference 
between the forward- and backward-evolving states \cite{Campagne112}. 

\begin{figure} [t]
\centering
\includegraphics[width=8.6cm]{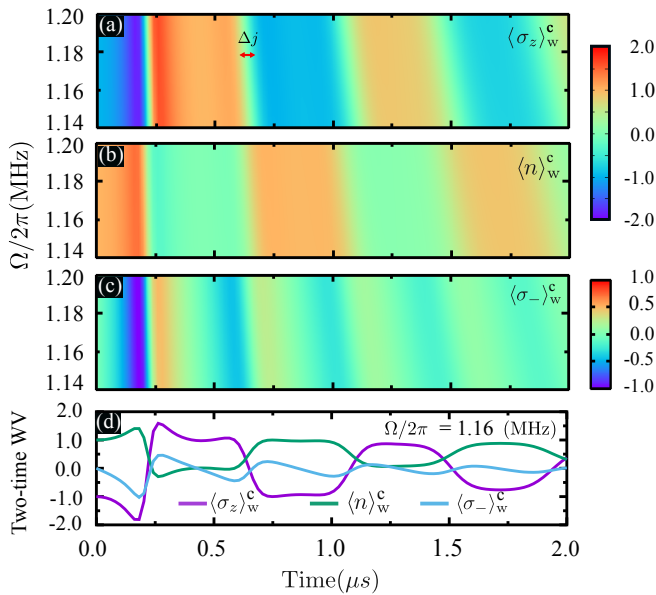}
\caption{
(Color online) Two-time correlation weak values of 
the atom population $\langle\sigma_z\rangle^{\rm c}_{\rm w}$ (a), 
the photon number $\langle n\rangle^{\rm c}_{\rm w}$ (b), 
and the fluorescence signal $\langle\sigma_-\rangle^{\rm c}_{\rm w}$ 
(c) in a wide range of frequency $\Omega$. 
(d) Extracted results at $\Omega/2\pi=1.16$ MHz.
}
\label{fig3}
\end{figure}

We next consider the two-time correlation weak values 
of these measured signals. Figure \ref{fig3}(a-c) show 
the two-time correlation weak values of the atom population 
$\langle\sigma_z\rangle^{\rm c}_{\rm w}$ (a), 
the photon number $\langle n\rangle^{\rm c}_{\rm w}$ (b), 
and the fluorescence signal $\langle\sigma_-\rangle^{\rm c}_{\rm w}$ 
(c) in a wide range of the frequency $\Omega$. 
Interestingly, these results do not behave Rabi oscillations. 
Indeed, the atom population exhibits a ``quantum jump" 
in between the ground state (blue areas) and the excited state(red areas). 
We define a ``jump-duration" ($\Delta j$) 
which is a necessary time interval 
for the atom state jumps from one state to another. 
This jump-duration increases in time for each fixed frequency. 
Along with this atom jumps, the photon number can be detected 
(red areas) or not (green areas,) respectively (Fig.\ref{fig3}(b).) 
Besides the jumping durations, the atom population 
and photon number tend to keep steady. 
In addition, the fluorescence signal in Fig. \ref{fig3}(c) 
shows a long decay between the two jumping durations. 
These effects can be seen clearly from the result in 
Fig. \ref{fig3}(d) extracted from the above Fig. \ref{fig3}(a-c) 
at the frequency $\Omega/2\pi = 1.16$ MHz. 
We also note that these results are obtained from  
Eq. \eqref{2weak_value_enlar} by using the enlarged quantum state 
$\varrho_t$. One of the advances of using this enlarged quantum state 
is that we can obtain dynamically the measured value at any time 
$0\le t\le T$ that no need to wait until the final time $T$. 
We found that the two-time correlation weak values also 
exhibit the amplification effect as can be seen from Fig. \ref{fig3}, 
where the measured signals excess outside the normal range [-1,1].
The results in Fig. \ref{fig3}(d) can apply to signal processing techniques. 
For example, one can  clearly distinguish between the two levels 
of the atom population signal (e.g., ``excited level" or ``ground level") 
because the signal keeps steady and therefore, its two-level can be triggered 
by setting a suitable threshold. Other kinds of signals, such as the photon number, 
the fluorescence signal, also can be explicitly detected.

\begin{figure} [t]
\centering
\includegraphics[width=8.6cm]{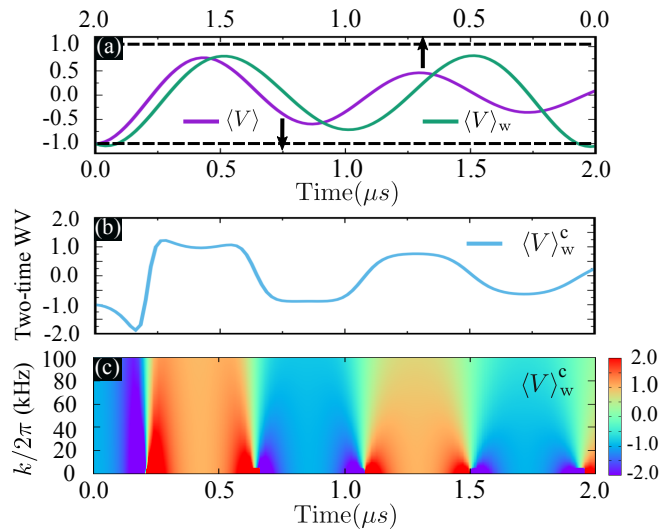}
\caption{
(Color online) (a) The average values conditioned on the past 
(only) or the future (only) of the voltage signals (dark-violet curve.) 
The bottom axis corresponds the average of 
the voltage signal conditioned on the past density matrix 
prepared in the ground state. The state propagates forward 
in time under the master equation Eq. \eqref{Lindblad_forward_R}. 
The top axis is the average of the voltage signal conditioned 
on the future density matrix prepared in the ground state. 
The state propagates backward in time by 
the master equation Eq. \eqref{Lindblad_backward_R}. 
The vine-green curve is the conventional weak value of 
the voltage signal corresponded to the quantum state prepared 
and post-selected on the ground state. 
It is obtained from the enlarged quantum state 
$\varrho_t$ as in Eq. \eqref{mean_V}. 
(b) The two-time weak value of the voltage signal at 
$k/2\pi = 95$ kHz. (c) The two-time weak value of 
the voltage signal for a wide range of $k$, 
which is obtained from Eq. \eqref{mean_2V}.
}
\label{fig4}
\end{figure}

As a second example, we apply our proposal to the case of 
weak continuously monitor a superconducting qubit as described in 
Refs. \cite{Foroozani116,Tan114}. The corresponding master equations 
are given by \cite{Foroozani116,Tan114}
\begin{align}
\dfrac{d\rho_t}{dt} &= -\dfrac{i\Omega}{2}[\sigma_y, \rho_t] 
+k(\sigma_z\rho_t\sigma_z-\rho_t), {\label{Lindblad_forward_R}}\\
\dfrac{dE_t}{dt} &= -\dfrac{i\Omega}{2}[\sigma_y, E_t] 
-k(\sigma_z E_t\sigma_z-E_t).{\label{Lindblad_backward_R}}
\end{align}
The enlarged master equation gives
\begin{align}\label{Lindblad_enlar_2}
\dfrac{d\varrho_t}{dt} = -\dfrac{i\Omega}{2}[\sigma_z\otimes \sigma_y, \varrho_t] + k[(\bm{I}_2\otimes\sigma_z)\varrho_t(\bm{I}_2\otimes\sigma_z)-\varrho_t].
\end{align}

Let us focus on the voltage signal in this case. 
We use the same $\Omega$ and $k$ as above, 
i.e., $\Omega/2\pi = 1.16$ MHz and $k/2\pi=95$ kHz. 
In Fig. \ref{fig4}(a), the dark-violet curve displays 
the voltage signal conditioned on a single preselected 
density matrix $\rho_0 = |g\rangle\langle g|$ 
or a single postselected density matrix $E_T=|g\rangle\langle g|$. 
These two values are the same in the time-reversal 
(we just see one curve because of the coincidence.) 
The result shows the damping effect as usual. 
Meanwhile, the vine-green curve in Fig. \ref{fig4}(a) 
exhibits the conventional weak value of the voltage signal, 
which is obtained from Eq. \eqref{mean_V}. 
Our study agrees with the previous study \cite{Foroozani116}. 
Figure \ref{fig4}(b) shows the jumping behavior 
of the two-time correlation weak value of the voltage signal 
for a fixed frequency at $k/2\pi = 95$ kHz. 
Notable, the jump duration $\Delta j$ increases in time 
at each fixed frequency. The effect even greater for a large range 
of $k$ as shown in Fig. \ref{fig4}(c).  
The signal amplification effect is again observed as in the previous example.

\section{Implementation}\label{secv}
In this section, we show that the $\mathcal{ES}$ 
can be implemented in some physical platforms. 
Assume that the $\mathcal{ES}$ is initially prepared 
in $\varrho_0$ at time $t=0$ and its evolution is provided 
by the von Neumann equation as 
$\varrho_t = \bm{\mathcal{U}}_t\varrho_0\bm{\mathcal{U}}_t^\dagger$, 
where $\bm{\mathcal{U}}_t =  {\rm exp}[-\frac{i}{\hbar}\bm{\mathcal{H}}t]$ 
and $\bm{\mathcal{H}}\equiv\sigma_z\otimes\bm{H}$ 
are the enlarged evolution and the enlarged Hamiltonian, respectively. 
The enlarged  evolution $\bm{\mathcal{U}}_t$ can be implemented 
by using entangling M\o lmer-S\o rensen gates $\bm{\mathcal{U}}_{\rm MS}$ 
as described in Ref. \cite{Ho97}. For example, to implement an $\mathcal{ES}$ 
that consists of one $\mathcal{OS}$ qubit and one ancillary qubit, 
we can prepare the initial enlarged state $\varrho_0$ 
in the bases of a four-level system \cite{Peterer114} 
in such way that it contains the pre- and 
postselected quantum states $\rho_0$ and $E_T$, 
respectively. The enlarged evolution is given by
\begin{align}\label{u}
 \bm{\mathcal{U}}_t =  {\rm exp}\Bigl[-\frac{i\Omega}{2}
 \bigl(\sigma_z\otimes\sigma_y\bigr)t\Bigr].
\end{align}
To implement $\bm{\mathcal{U}}_t$, we need 
(i) apply the M\o lmer-S\o rensen gate onto both 
the system qubit and the ancillary qubit, 
(ii) apply a local single-qubit rotation
onto the ancillary qubit, and 
(iii) apply the M\o lmer-S\o rensen again.
Following Ref. \cite{Casanova108}, 
we can implement $\bm{\mathcal{U}}_t$ by
\begin{align}{\label{operator_App_2}}
\bm{\mathcal{U}}_t = \bm{\mathcal{U}}_{\rm MS}
(-\dfrac{\pi}{2},\dfrac{\pi}{2})e^{i\frac{\Omega t}{2}\sigma_x}
\bm{\mathcal{U}}_{\rm MS}^\dagger(-\dfrac{\pi}{2},\dfrac{\pi}{2})\;.
\end{align}

For the second example described in Sec. \ref{seciv}, 
the ancillary qubit is kept out of the environment and 
freely evolves under the system Hamiltonian 
$\bm{H} \equiv \sigma_z$. We can implement the $\mathcal{ES}$ 
in a pair of qubits \cite{Majer449}, where one qubit is driven 
at resonance with the cavity while the other is kept out of the resonance. 
We then also apply a sequence of the entangling 
M\o lmer-S\o rensen gates as we have already described above. 
The $\mathcal{ES}$ thus can be implemented in 
superconducting circuits and also other physical systems, 
such as ion-traps \cite{Casanova1,Leibfried75,Batt2252,Lamata19}, 
quantum photonics \cite{Loredo116}. In superconducting circuits, 
for example, one qubit is coupled to a cavity and plays the role of the 
$\mathcal{OS}$, while the remain qubit can be viewed as 
the ancillary qubit out of the cavity. Similarly, in ion-traps, 
one trapped ion can be addressed as the $\mathcal{OS}$ 
resonance at laser frequency, while another trapped ion is 
the ancillary qubit and does not interact with the laser. 
In all cases, the ancillary qubit can be verified by changing 
the two-level energy splitting, using a magnetic field crossing 
the qubit \cite{Para175}.   
 
\section{Discussions and Conclusions}\label{secvi}

We remind that our proposal about the enlarged system 
($\mathcal{ES}$) in the enlarged Hilbert space 
plays the role of a quantum simulator, 
which is a one-to-one mapping between 
an original quantum system (the simulated system, $\mathcal{OS}$) 
to a given mathematical model (the simulator system,) 
which is more controllable for reproducing 
the dynamics of the quantum system 
\cite{Feynman21,Georgescu86}. 
The primary task of quantum simulators is 
to solve the dynamical time-dependent 
Schr\"{o}dinger equation by fundamental laws 
of nature and also can be demonstrated 
in many physical models \cite{Georgescu86}.
Specifically, in our case, the simulator system is 
the $\mathcal{ES}$ where it simulates 
the forward and backward evolutions of the $\mathcal{OS}$. 
As a consequence, by continuous probing of the 
$\mathcal{ES}$, we also can control and gain 
the information in the $\mathcal{OS}$.

In conclusion, we found that the measured signals 
of an $\mathcal{OS}$ bounded by past and future conditions 
can be monitored continuously in an $\mathcal{ES}$. 
Specifically, the weak value in the $\mathcal{ES}$ 
is two-time correlated, which we name as 
the two-time correlation weak value. 
We showed that the two-time correlation weak value 
can be obtained dynamically at a given time 
by tracking the trajectory of the enlarged state. 
We have applied our proposal to the concept of 
a superconducting qubit driven by a laser field at 
the resonance frequency and shown the quantum jump effect 
in the measured signals.
We have also observed the amplification effect of the two-time correlation weak value 
as well as its application in the signal processing techniques. 
Our proposal thus provides significant benefits in scientific and technological
applications.
It also can motivate and guide 
further various exciting experiments.

\begin{acknowledgements}
We would like to thank Y. Kondo of Kindai University 
for useful discussions. 
This work was supported by JSPS KAKENHI Grant Number JP19K14620.
\end{acknowledgements}

\begin{widetext}
\appendix
\section{Master equation for an enlarged density matrix}\label{appA}
\setcounter{equation}{0}
\renewcommand{\theequation}{A.\arabic{equation}}

The forward-evolving state propagates forward in 
time from the initial state $\rho_0$ by the master equation
\begin{align}{\label{Lindblad_forward_appA}}
\dfrac{d\rho_t}{dt} = -\dfrac{i}{\hbar}[\bm{H}, \rho_t] 
+\sum_n\dfrac{1}{2}\Bigl[2\bm{C}_n\rho_t\bm{C}_n^\dagger 
- \{\bm{C}_n^\dagger \bm{C}_n,\rho_t\}\Bigr],
\end{align}
and the backward-evolving state propagates backward 
in time from the final state $E_T$ by the master equation
\begin{align}\label{Lindblad_back_appA}
\dfrac{dE_t}{dt} = -\dfrac{i}{\hbar}[\bm{H}, E_t] 
-\sum_n\dfrac{1}{2}\Bigl[2\bm{C}_n^\dagger E_t\bm{C}_n 
- \{\bm{C}_n^\dagger \bm{C}_n, E_t\}\Bigr].
\end{align}
Because of the time symmetry in the measurement interval 
$t\in[0,T]$, we introduce an alternative ``time-forward version" 
of the  backward master equation Eq. \eqref{Lindblad_back_appA}, 
which evolves forward in time as 
\begin{align}\label{Lindblad_back_for_appA}
\dfrac{dE_{T-t}}{dt} &= \dfrac{i}{\hbar}[\bm{H}, E_{T-t}] 
+\sum_n\dfrac{1}{2}\Bigl[2\bm{C}_n^\dagger E_{T-t}\bm{C}_n 
- \{\bm{C}_n^\dagger \bm{C}_n, E_{T-t}\}\Bigr]. 
\end{align}
For $t=0$, the initial state is prepared by $E_T$, 
which corresponds to the postselection state. 
In other words, the postselection state is prepared at the beginning, 
in a similar manner to the previous work \cite{Ho97}. 
The solution of this equation at time $t$ corresponds to $E_{T-t}$, 
which is also the solution of Eq \eqref{Lindblad_back_appA} at time $T-t$.

We next define the enlarged quantum state as
\begin{align}{\label{varrho_appA}} 
\varrho_t \equiv
\begin{pmatrix}
 [\varrho_t]^{00} & [\varrho_t]^{01} \\
  [\varrho_t]^{10} & [\varrho_t]^{11} 
 \end{pmatrix}
 =\dfrac{1}{2}
\begin{pmatrix}
  \rho_t & 0_d \\
  0_d & E_{T-t} 
 \end{pmatrix}\;,
\end{align}
where $[\varrho_t]^{ij}$ are $d\times d$ bock matrices.
The combination of two forward master equations 
\eqref{Lindblad_forward_appA} and 
\eqref{Lindblad_back_for_appA} 
in the following way will give the enlarged master equation
%
\begin{align}{\label{Enlar_mas_appA}} 
\dfrac{d\varrho_t}{dt} = -\dfrac{i}{\hbar}\Biggl[
\begin{pmatrix}
  \bm{H} & 0_d \\
  0_d &-\bm{H} 
 \end{pmatrix},\varrho_t\Biggr]
 +\sum_n\dfrac{1}{2}\Biggl[
2\begin{pmatrix}
  \bm{C}_n & 0_d \\
  0_d &\bm{C}_n^\dagger 
 \end{pmatrix}
 \varrho_t
  \begin{pmatrix}
  \bm{C}_n^\dagger & 0_d \\
  0_d &\bm{C}_n 
 \end{pmatrix}
 -\Biggl\{
  \begin{pmatrix}
  \bm{C}_n^\dagger & 0_d \\
  0_d &\bm{C}_n^\dagger
 \end{pmatrix}
   \begin{pmatrix}
  \bm{C}_n & 0_d \\
  0_d &\bm{C}_n
 \end{pmatrix},
 \varrho_t\Biggr\}
 \Biggr].
\end{align}
%
We set the enlarged operators  
$\bm{\mathcal{H}}\equiv\sigma_z\otimes\bm{H}$, 
$\bm{\mathcal{C}}\equiv|0\rangle\langle 0|\otimes\bm{C}
+|1\rangle\langle 1|\otimes\bm{C}^\dagger$, 
and $\bm{\bar\mathcal{C}}\equiv \bm{I}_2\otimes\bm{C}$, 
the enlarged master equation will give 
Eq. \eqref{Lindblad_enlar} in the main text.

For qubits case, we plot in Fig. \ref{fig5} 
the quantum trajectories of the forward-evolving (left), 
the backward-evolving (right) and the enlarged (middle) 
quantum states in the $x-z$ planes of the Bloch spheres against time. 
We consider here the resonance fluorescence case 
as in the main text with the pre and postselected states 
are given in the ground state. The red arrows indicate the direction 
of time evolution starting from $t = 0$. To illustrate the enlarged state, 
we divide it into block matrices $[\varrho_t]^{ij}$ as shown in the figure. 
More precisely, we have $[\varrho_t]^{00}=\rho_t/2$, $[\varrho_t]^{01}
=[\varrho_t]^{10}=0_d$ ($d=2$ for the qubit case.) 
Specifically, $[\varrho_t]^{11}=E_{T-t}/2$, 
which is the time-reversal of $E_t/2$ in the interval $[0,T]$. 

\begin{figure} [t]
\centering
\includegraphics[width=14cm]{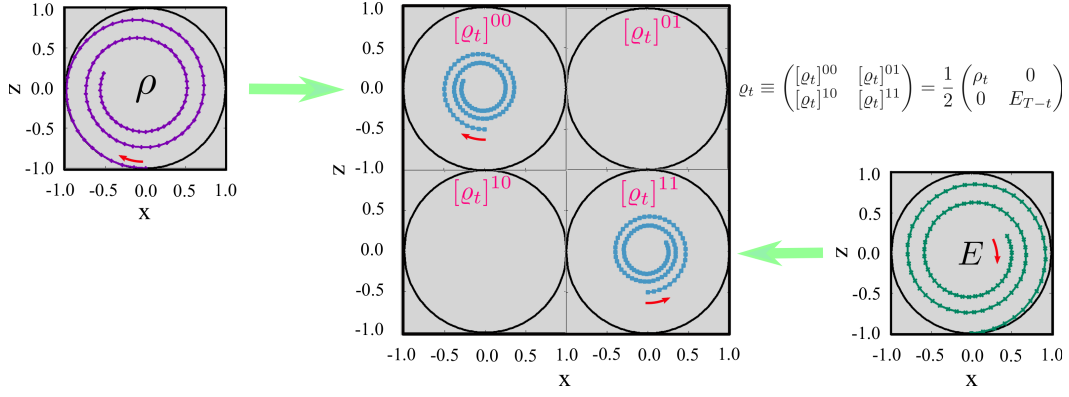}
\caption{
(Color online) The quantum trajectories of 
the forward-evolving (left), the backward-evolving (right) 
and the enlarged (middle) quantum states in the $x-z$ planes 
of the Bloch spheres against time. 
The red arrows indicate the direction of time evolution starting from $t = 0$.
}
\label{fig5}
\end{figure}

\section{Analytical solution for the voltage signal in the enlarged system}
\label{appB}
\setcounter{equation}{0}
\renewcommand{\theequation}{B.\arabic{equation}}

Let us first decode the forward- and the backward-evolving states as follows 

\begin{align}\label{past_appB}
\notag\rho_t &= 2\mathcal{M}\varrho_t\mathcal{N}\\
\notag&=2
\begin{pmatrix}
 1 & 0 & 1 & 0 \\
0 & 1&0&1
 \end{pmatrix}
 \begin{pmatrix}
 \varrho^{00}_t & \varrho^{01}_t & \varrho^{02}_t &\varrho^{03}_t \\
 \varrho^{10}_t & \varrho^{11}_t & \varrho^{12}_t &\varrho^{13}_t \\
  \varrho^{20}_t & \varrho^{21}_t & \varrho^{22}_t &\varrho^{23}_t \\
   \varrho^{30}_t & \varrho^{31}_t & \varrho^{32}_t &\varrho^{33}_t 
 \end{pmatrix}
 \begin{pmatrix}
 1 & 0  \\
0 & 1\\
0& 0\\
0&0
 \end{pmatrix}\\
 &=2 \begin{pmatrix}
\varrho^{00}_t & \varrho^{01}_t  \\
\varrho^{10}_t  & \varrho^{11}_t \end{pmatrix},
\end{align}
where we have used the block matrices  
$[\varrho_t]^{01}=[\varrho_t]^{10}=0_d$. 
Similarly, we also have 
\begin{align}\label{future_appB}
 E_t &= 2\mathcal{M}\varrho_{T-t}(\sigma_x
 \otimes\bm{I}_n)\mathcal{N}=2 \begin{pmatrix}
\varrho^{22}_{T-t} & \varrho^{23}_{T-t}  \\
\varrho^{32}_{T-t}  & \varrho^{33}_{T-t} \end{pmatrix}.
\end{align}
We now insert these two equations into 
the probability of the outcome of $V_t$ 
of the continues measurement, which is defined by 
\cite{Gammelmark111}

\begin{align}\label{PV_appB}
 \notag P(V) &= \dfrac{{\rm Tr}(\Omega_V\rho_t\Omega_V^\dagger E_{t})}
 {\sum_V {\rm Tr}(\Omega_V\rho_t\Omega_V^\dagger E_{t})} \\
 &\propto \varrho^{00}_t\varrho^{22}_{T-t}e^{-(V-1)^2/2a^2}
+\varrho^{11}_t\varrho^{33}_{T-t}e^{-(V+1)^2/2a^2} 
+(\varrho^{10}_t\varrho^{23}_{T-t}
+\varrho^{01}_t\varrho^{32}_{T-t})e^{-(V^2+1)/2a^2}.
\end{align}

Taking the integral over the outcome $V$, 
i.e., $\int P(V)V dV$, the conventional weak value of the voltage signal is given by
\begin{align}\label{mean_V_appB}
\langle V\rangle_{\rm w} =  \dfrac{\varrho^{00}_t\varrho^{22}_{T-t}
-\varrho^{11}_t\varrho^{33}_{T-t}}{\varrho^{00}_t\varrho^{22}_{T-t}
+\varrho^{11}_t\varrho^{33}_{T-t}+\varrho^{10}_t\varrho^{23}_{T-t}
+\varrho^{01}_t\varrho^{32}_{T-t}}.
\end{align}
This conventional weak value can be measured continuously, 
however, the forward-evolving and backward-evolving states 
are obtained separately as experimentally performed in 
Ref. \cite{Tan114,Foroozani116}. 
Even in our proposal of enlarged Hilbert space in this work, 
the conventional weak value only is acquired whenever the trajectories 
at time $t$ and $T-t$ are given. 
Nevertheless, for the two-time correlation weak value, 
the voltage signal gives
\begin{align}\label{mean_2V_appB}
\langle V\rangle^{\rm c}_{\rm w} =
\dfrac{\varrho^{00}_t\varrho^{22}_t
-\varrho^{11}_t\varrho^{33}_t}{\varrho^{00}_t\varrho^{22}_t
+\varrho^{11}_t\varrho^{33}_t+\varrho^{10}_t\varrho^{23}_t
+\varrho^{01}_t\varrho^{32}_t}.
\end{align}
In this form, the two-time correlation weak value 
can be obtained dynamically by tomography 
the enlarged quantum state $\varrho_t$ at time $t$, 
which is an advantage of our proposal for the ``two-timer." 
\end{widetext}



\end{document}